# Zero-shot Voice Conversion via Self-supervised Prosody Representation Learning

Shijun Wang
AIML Lab, School of Computer Science
University of St. Gallen
St. Gallen, Switzerland
shijun.wang@unisg.ch

Damian Borth
AIML Lab, School of Computer Science
University of St. Gallen
St. Gallen, Switzerland
damian.borth@unisg.ch

*Abstract*—Voice Conversion (VC) for unseen speakers, also known as zero-shot VC, is an attractive research topic as it enables a range of applications like voice customizing, animation production, and others. Recent work in this area made progress with disentanglement methods that separate utterance content and speaker characteristics from speech audio recordings. However, many of these methods are subject to the leakage of prosody (e.g., pitch, volume), causing the speaker voice in the synthesized speech to be different from the desired target speakers. To prevent this issue, we propose a novel self-supervised approach that effectively learns disentangled pitch and volume representations that can represent the prosody styles of different speakers. We then use the learned prosodic representations as conditional information to train and enhance our VC model for zero-shot conversion. In our experiments, we show that our prosody representations are disentangled and rich in prosody information. Moreover, we demonstrate that the addition of our prosody representations improves our VC performance and surpasses state-of-the-art zero-shot VC performances.

## I. INTRODUCTION

Voice Conversion (VC) converts the voice of a source speech to the voice of a target speech while maintaining the source's content. VC is widely utilized in a variety of applications, including privacy protection, the entertainment sector, and many more.

Traditional VC approaches like Gaussian mixture models [1] and restricted Boltzmann machines [2] require parallel data, i.e., different speakers need to read the same contents. Such datasets are hard to obtain. Recently, VC models that can be trained on non-parallel data have been proposed and show promising results. Generative models like Variational autoencoders (VAEs) [3], [4], Generative Adversarial Network (GAN) [5], [6] and Flow [7] are widely studied for nonparallel VC. Although these models can generate speech with good quality, they are limited to VC among speakers from the training dataset and struggle to synthesize a voice from previously unseen speakers (zero-shot voice conversion).

Zero-shot voice conversion is a new research direction that focuses on conversion between speakers that are unseen during training, and has become more and more popular since it can address the real-world problem of speaker information that is sometimes inaccessible or unknown during training.

One common zero-shot VC approach is to disentangle spoken contents and speaker characteristic information from speech audios [8]–[10]. With this method, unseen speaker characteristic information can be extracted from the speech, and can be recombined with content information from other speech.

Despite the efforts on disentanglement in such models, the prosody leakage issue might happen, where prosody elements like pitch or volume are partially damaged or even totally lost during conversion. Prosody leakage results in the phenomenon that the speaker voices in the synthesized speech samples are different from the voices of the desired target speakers. The prosody leakage is often caused by the self-reconstruction training loss in most of these related models [11], which expects the speakers in the input and output speech to be the same. Under the self-reconstruction loss, the model is likely to encode speaker prosody information in the latent representations (e.g., content representations). However, because the source and target speakers are different during inference, the source prosody information that resides in the latent representation might hinder conversion.

If we explicitly provide the prosody information of the target speaker, the model might obtain benefits and avoid prosody leakage. Thus, in this paper, we propose a novel self-supervised approach to effectively learn pitch and volume representations, two crucial prosody elements [12], from diverse speakers. Such representations represent speakers' prosody style (e.g, a speaker tends to speak in a high pitch or a low volume). The extracted prosodic representations are then used as useful conditional information to train and enhance our zero-shot VC model. The addition of prosody representation is similar to the addition of speaker embedding X-vectors [13], except that prosody representations provide speaker prosody information rather than speaker characteristic information.

We summarize our contributions as follows: (i) We propose a novel zero-shot VC approach. (ii) We introduce a self-supervised method to extract disentangled prosody representations without the requirement of labels or hand-crafted prosody features. (iii) Our experiments show that the prosody encoder can learn disentangled prosody representations from diverse speakers. Moreover, we demonstrate that our VC model obtains improvements by adding the prosody representations and outperforms state-of-the-art zero-shot VC models.

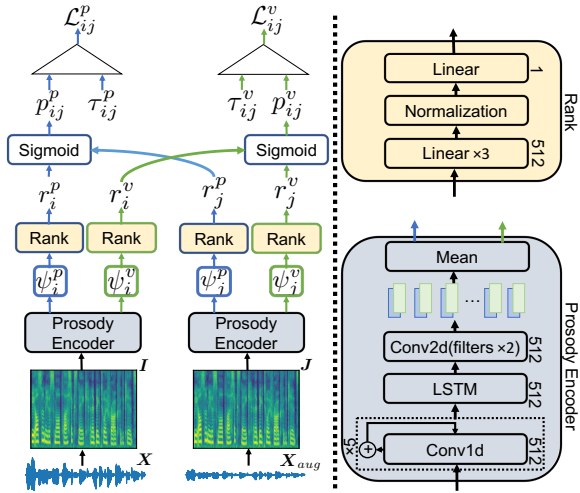

Fig. 1. The left is the training pipeline of our prosody encoder, the right is the architecture of our prosody encoder and Rank module. In the pipeline, we first convert a speech waveform pair ($X$, $X_{aug}$) into a Mel-Spectrogram pair ($I$, $J$) as the training input. Our prosody encoder extracts pitch and volume representation $\psi^p$ and $\psi^v$ from each Mel-Spectrogram. Pitch Rank module (blue frame) and volume Rank module (green frame) produce pitch score $r^p$ and volume score $r^v$ based on the $\psi^p$ and $\psi^v$, respectively. Two scores of the same prosody type but from different inputs are then sent to the Sigmoid function to get $p_{ij}$. $p_{ij}$ indicates the probability that $X$ is ranked higher than $X_{aug}$ regarding pitch/volume. Lastly, we calculate cross entropy loss $\mathcal{L}_{ij}$ from $p_{ij}$ and the augmentation hyper-parameter $\tau_{ij}$.

## II. RELATED WORKS

In this section, we summarize related works on zero-shot voice conversion and speech prosody disentanglement.

### A. Zero-shot Voice Conversion

Zero-shot voice conversion focuses on the conversion between the speakers who are unseen in the training dataset. Disentangling spoken content and speaker characteristics from recording speech data is a popular strategy for this task. With disentanglement, the content from the source speech and the speaker characteristic from the target can then be combined during conversion to achieve zero-shot VC. AutoVC [8] applies a pre-trained speaker encoder and a vanilla autoencoder with a carefully tuned bottleneck to enable disentanglement. Authors in [9] use Instance Normalization (IN) to separate linguistic content and speaker characteristics. [10] takes into account different activation functions for better content-speaker disentanglement. In [14], [15], the authors use Vector Quantization (VQ) [16] to remove speaker information from the content. However, prosody leakage issue has been found in some of these related works. One cause [11] is that these models' training loss is based on self-reconstruction, which expects the speakers in the input and output speech to be the same. Hence the model is likely to encode speaker prosody information in the latent representations (e.g., content representations). However, because the source and target speakers are different during inference, the source prosody information encoded in the latent representations might impede conversion.

The authors in [11], [17] mitigate prosody leakage issue by conditioning on the pitch prosody (F0) information. In conclusion, effective learning of prosody is critical for better representing unseen speakers in the zero-shot VC task.

### B. Speech Prosody Disentanglement

Speech prosody disentanglement has been studied for many years and still remains an important topic due to its application in controllable or emotional speech synthesis. Text-to-Speech models like [18], [19] can learn disentangled prosody representations and synthesize a speech with controllable prosody. However, these models require hand-crafted prosody features. The model in [20] can disentangle speech prosody elements, although hand-crafted prosody features are still required. The model in [21] can learn prosody representations without the need for hand-crafted features, but there is little empirical evidence that learned representations are disentangled. Self-supervised learning has recently received a lot of attention in a variety of domains due to its ability to learn useful representations without the need of labels [22]–[24]. Self-supervised methods learn representations by taking advantage of the data invariance or variance that augmentation transforms provide [25]–[27]. However, there are very few works investigating to use self-supervised methods to learn prosody representation. In this paper, inspired by RankNet [28], where using rank labels to learn visual attribute representations, we use self-supervised methods and learn prosody representations based on the prosody variance provided by augmentations.

## III. METHOD

In this section, we introduce our data augmentation, prosody encoder model and describe our zero-shot voice conversion model.

### A. Data Augmentation

Since we learn prosody representations based on the prosody variance that augmentation transforms provide, we first need to describe our augmentation process. We denote speech waveform data as $X$, and apply two augmentation functions $X_{aug} = \text{Transform}_{pro}(X; \tau)$ for pitch stretch or volume adjustment, where $pro \in \{p, v\}$ is defined as the prosody set and $p$ stands for pitch, $v$ stands for volume. The hyper-parameter $\tau \in (0, 1)$ indicates the augmentation intensity. Given a specific augmentation function, $\tau < 0.5$ means negative augmentations (decrease pitch/volume), while $\tau > 0.5$ means positive augmentations (increase pitch/volume). Lastly, $\tau = 0.5$, indicates no augmentation is applied.

### B. Prosody Encoder

Inspired by RankNet [28], we use a pair of inputs to train our prosody encoder, one of which is different from the other in terms of pitch or volume information. Instead of requiring ground truth ranked pairs in the original RankNet work, we apply augmentation transforms to construct the training pairs. Additionally, different from RankNet, which requires expert rank labels, we consider the augmentation hyper-parameter $\tau$ our rank label.

We use the pair ($X$, $X_{aug}$) as the training input for the prosody encoder. The pipeline is illustrated in Fig. 1. We first convert the audio pair into a pair of Mel-Spectrograms ($I$, $J$). Then, we feed the Mel-Spectrogram pair into the prosody encoder. For each Mel-Spectrogram, the prosody encoder produces two 512D prosody representations $\psi^p$ and $\psi^v$, for pitch and volume information respectively. Finally, we apply two Rank modules to respectively rank pitch and volume prosody representations by mapping $\psi$ into $r \in \mathbb{R}^1$. $r$ is used as a score of the prosody intensity (large $r$ value, high pitch/volume; small $r$ value, low pitch/volume).

With the output $r$ of our Rank modules, we define the training loss to achieve disentangled prosody representations:

$$\mathcal{L}_{\text{prosody}} = \sum_{pro \in \{p,v\}} \mathcal{L}_{ij}^{pro}, \text{ where} \quad (1)$$

$$\mathcal{L}_{ij}^{pro} = -\tau_{ij}^{pro}\log(p_{ij}^{pro}) - (1-\tau_{ij}^{pro})\log(1-p_{ij}^{pro}), \quad (2)$$

$$p_{ij}^{pro} = \frac{1}{1+e^{-(r_i^{pro}-r_j^{pro})}}. \quad (3)$$

The idea behind the training loss $\mathcal{L}_{\text{prosody}}$ is to learn prosody representations by predicting the augmentation intensity $\tau$. We first calculate the difference between prosody scores $r_i^{pro}$ and $r_j^{pro}$ from the original and augmented inputs, then we send the score difference $r_i^{pro} - r_j^{pro}$ into a Sigmoid function to get $p_{ij}^{pro}$ (Equation (3)), $p_{ij}^{pro}$ indicates the probability that $X$ is ranked higher then $X_{aug}$ regarding prosody $pro$. We then calculate the cross entropy loss between $p_{ij}^{pro}$ and the augmentation intensity hyper-parameter $\tau^{pro}$ (Equation (2)). Since we do two types of augmentations (prosody stretch and volume adjustment), we sum $\mathcal{L}_{ij}^p$ (pitch) and $\mathcal{L}_{ij}^v$ (volume) up to produce the overall loss $\mathcal{L}_{prosody}$ (Equation (1)).

Furthermore, to achieve the disentanglement between pitch representation $\psi^p$ and volume representation $\psi^v$, we only apply one type of augmentation transform (either pitch stretch or volume adjustment) for each training iteration. The motivation is to keep one type of prosody representations similar while encouraging the dissimilar of the representations of the other prosody element. For instance, in one iteration, we perform pitch stretch and obtain $X_{aug}$ from augmentation function $\text{Transform}_{pitch}(X;\tau^p = 0.3)$ (decrease the pitch). Since the volumes of $X$ and $X_{aug}$ are not expected to be affected and remain the same, we set $\tau^v = 0.5$ (no volume augmentation is applied). Thus, when we optimize the prosody encoder with loss $\mathcal{L}_{ij}^v$ (Equation (2)), $p_{ij}^v$ should be close to 0.5, which means the volume score $r_i^v$ of $X$ and the volume score $r_j^v$ of $X_{aug}$ should be close. In other words, we encourage our prosody encoder produce similar volume representations $\psi_i^v$ and $\psi_j^v$. On the other side, pitch representations $\psi_i^p$ and $\psi_j^p$ are expected to be dissimilar because of the pitch augmentation. By doing this, we enable our prosody encoder to output two disentangled representations, which are aiming to represent two different types of prosody elements (pitch and volume).

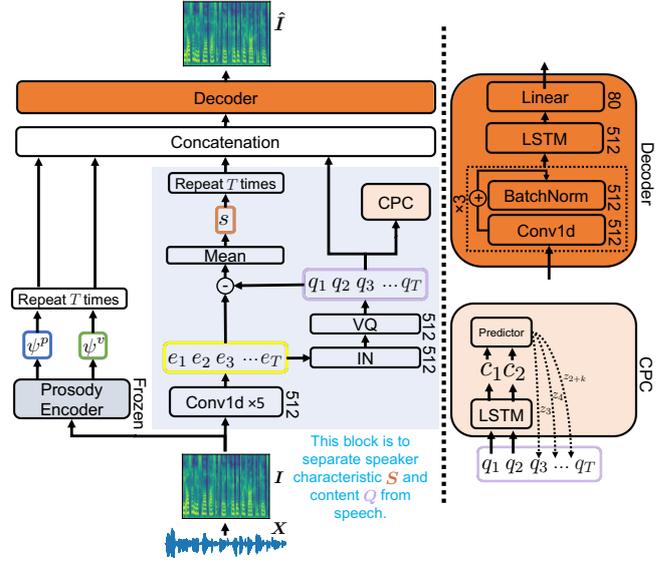

Fig. 2. The left is the training pipeline of our zero-shot voice conversion model, the right is the architecture of our decoder and Contrastive Predictive Coding (CPC). In the pipeline, we first convert the waveform $X$ into Mel-Spectrogram $I$. We send $I$ into two paths, the left path is to extract pitch and volume representations $\psi^p$ and $\psi^v$ with a pre-trained and frozen prosody encoder. The right path (the big blue block) is to produce content information $Q$ and speaker characteristic $S$. We first use convolutional layers to process $I$ and get $E$. Then we employ Instance Normalization (IN), Vector Quantization (VQ) and CPC to screen out content $Q$ from $E$. Speaker characteristics can be obtained with $S = \mathbb{E}||E - Q||$. Lastly, with $\psi^p$, $\psi^v$, $Q$ and $S$, we use a decoder to reconstruct the input $I$.

### C. Zero-shot Voice Conversion Model

Like many other works, the training of our VC model is based on the self-construction loss, and additional effective methods are applied to yield disentangled spoken content and speaker characteristic information. But the key difference is that we train the VC model with pitch and volume representations $\psi^p$ and $\psi^v$ from our prosody encoder as conditional information. The pipeline is showed in Fig. 2.

We first convert the audio waveform $X$ into Mel-Spectrogram $I$ and send it into two paths. The left path includes a prosody encoder, which is pre-trained and remains frozen during the training of our VC model. As discussed in Sec. III-B, our prosody encoder produces pitch and volume representations $\psi^p$ and $\psi^v$ from speech. For the right path (the big blue block), our target is to separate spoken contents and speaker characteristic information from the speech.

Specifically, on the right path, we first feed Mel-Spectrogram $I$ into 5 layers of 1D convolutions with ReLU activation to produce latent representation $E = \{e_1, e_2, ..., e_T\}$, where $e_t \in \mathbb{R}^{512}$, and $T$ is the length of the Mel-Spectrogram.

To remove acoustic information (e.g., speaker characteristics, prosody information) from $E$, and screen out spoken content information $Q = \{q_1, q_2, ..., q_T\}$, where $q_t \in \mathbb{R}^{512}$, we adopt Instance Normalization (IN) [9], Vector Quantization (VQ) [14] and Contrastive Predictive Coding (CPC) [29] into our VC model. IN can remove invariant information (like the speaker characteristic) away. VQ extracts content information by mapping similar vectors into one vector,

$$q_t = \arg\min_{q^* \in \mathcal{Q}^V}(\|e_t - q^*\|_2^2), \quad (4)$$

where $\mathcal{Q}^V$ is a learnable discrete codebook with limited numbers of vectors (we denote the number of discrete vectors in codebook as codebook size $V$). VQ takes $e_t$ and selects the closest $q$ from a codebook based on the Euclidean distance. We train our VC model and update the codebook $\mathcal{Q}^V$ with

$$\mathcal{L}_{vq} = \|\text{sg}[\boldsymbol{E}] - \boldsymbol{Q}\|_2^2 + \beta\|\boldsymbol{E} - \text{sg}[\boldsymbol{Q}]\|_2^2, \quad (5)$$

where sg[·] denotes the stop-gradient operator. The first term $\mathcal{L}_{vq}$ updates the codebook $\mathcal{Q}^V$, while the second term prevents $\boldsymbol{E}$ from growing arbitrarily, hyper-parameter $\beta$ is the weight between these two terms.

CPC is another effective way to extract content information. It applies an InfoNCE loss [30] to predict the positive representation from a set of negative representations. Specifically, we pass $\boldsymbol{Q}$ through a LSTM to produce context $\boldsymbol{C} = \{c_1, c_2, ..., c_T\}$ and we build linear predictors $Pred^k$ ($0 < k \leq K$) that take $c_t$ and output a vector $z_{t+k}$. We define our CPC loss as

$$\mathcal{L}_{cpc} = \frac{1}{K}\sum_{k=1}^{K}\log\frac{\exp(\text{dot}(Pred^k(c_t), q_{t+k}))}{\sum_{n \in \mathcal{N}_q}\exp(\text{dot}(Pred^k(c_t), q_n))}. \quad (6)$$

We enforce similarity by minimizing the dot product (dot(·, ·)) between prediction $z_{t+k}$ and the positive vector $q_{t+k}$ while account for dissimilarity by maximizing the dot product between $z_{t+k}$ and the negative vectors $\mathcal{N}_q$ that are randomly drawn from our training batch.

Afterwards, with content information $\boldsymbol{Q}$, we follow [14] and produce speaker characteristic $\boldsymbol{S} = \mathbb{E}[\boldsymbol{E} - \boldsymbol{Q}]$.

To prepare the input of the decoder, we repeat pitch representation $\psi^p$ and volume representation $\psi^v$ for $T$ times in order to concatenate them with content representation $\boldsymbol{Q}$. The decoder utilizes prosody representations $\psi^p$ and $\psi^v$, content information $\boldsymbol{Q}$ and speaker characteristic $\boldsymbol{S}$ to generate the Mel-Spectrogram $\hat{\boldsymbol{I}}$. We use the reconstruction loss

$$\mathcal{L}_{rec} = \|\boldsymbol{I} - \hat{\boldsymbol{I}}\|_1^1 + \|\boldsymbol{I} - \hat{\boldsymbol{I}}\|_2^2. \quad (7)$$

Both L1 ($\|\cdot\|_1$) and L2 ($\|\cdot\|_2$) distances are applied to increase the stability of the training process.

Lastly, to train our VC model, we define loss $\mathcal{L}_{vc}$ as

$$\mathcal{L}_{vc} = \mathcal{L}_{rec} + \mathcal{L}_{vq} + \mathcal{L}_{cpc}. \quad (8)$$

### D. Inference

For voice conversion, we have source speech and target speech, and we expect to generate a speech with content from the source, while speaker characteristic and prosody from the target. Thus, during our inference phase, we obtain content $\boldsymbol{Q}$ from the source speech (the right path in Fig. 2), pitch and volume prosody representations $\psi^p$ and $\psi^v$ from target speech, and speaker characteristic information $\boldsymbol{S}$ from the target speech as well (the right path in Fig. 2). Lastly, we synthesize a new speech with the decoder from the aforementioned information.

## IV. EXPERIMENTS

We devote this section to our empirical evaluation. Synthesized samples can be found in our demo page[1].

### A. Dataset

We conduct experiments on the VCTK dataset [31], which is a popular dataset for the VC task. VCTK includes 109 English speakers, each speaker reads about 400 sentences. After preprocessing, we remove one speaker due to the insufficient samples of this speaker. We then split the remaining 108 speakers into training and testing data sets. There are 88 speakers in the training set, which we refer to *seen speakers*, the left 20 speakers in the testing set are *unseen speakers*. Additionally, in VCTK, 91 out of 109 speakers in VCTK read "Please call Stella.", which we utilize in one of the experiments (Sec. IV-C2).

### B. Implementation Detail and Experiment Setup

For augmentaion, we use the library *pysox*[2] to augment waveform files. During data preprocessing, we downsample all VCTK waveform files from 48000Hz to 22050Hz, and convert these files into 80-bin Mel-Spectrograms with 1024 STFT window size and 5.8 milliseconds hop size.

We train our models on the *seen speakers* set, and evaluate the models on the *unseen speakers* set. After fine-tuning, we set $\beta$ in Equation (5) to 0.25, and $k$ in Equation (6) to 20. We train our prosody encoder with a batch size of 32 and a learning rate of 0.0001 for 150k iterations. We train the VC model for 300k iterations, and set the learning rate to 0.00003, the batch size to 64. We use the Adam optimizer for both models.

Since our VC model outputs Mel-Spectrograms, we need to convert them into waveform files. Thus, we employ a pre-trained vocoder PWGAN [32] to perform this task.

We include two state-of-the-art VC models as our baselines, AGAIN [10] and VQVC+ [15]. AGAIN can do zero-shot VC by Activation Guidance and Adaptive Instance Normalization. VQVC+ is a U-net like VC model with hierarchical VQ layers. To make fair comparisons, all baseline models are trained with the same dataset. Additionally, in the following experiments, we use the term **Ours** to denote out zero-shot VC model that is conditioned on the pitch and volume prosody representations.

### C. Study of the Prosody Encoder

In this section, we carry out experiments to see if our proposed prosody encoder can produce pitch and volume representations properly. Besides, we want to investigate whether the prosody encoder is able to extract distinct prosody representations for different unseen speakers.

---

[1] https://anonymous.4open.science/w/IJCNN2022_demo-5003/
[2] https://pysox.readthedocs.io/en/latest/

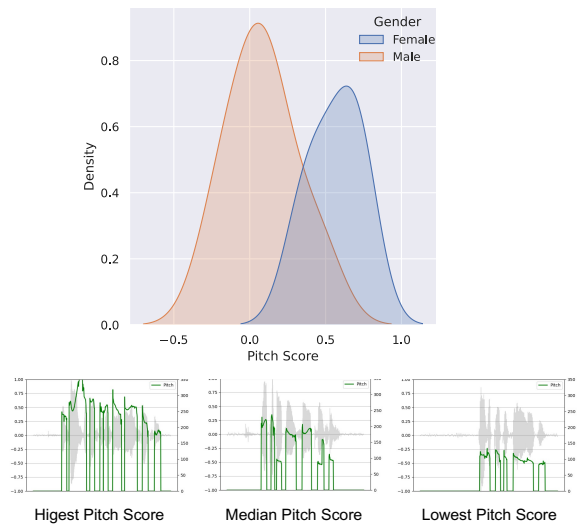

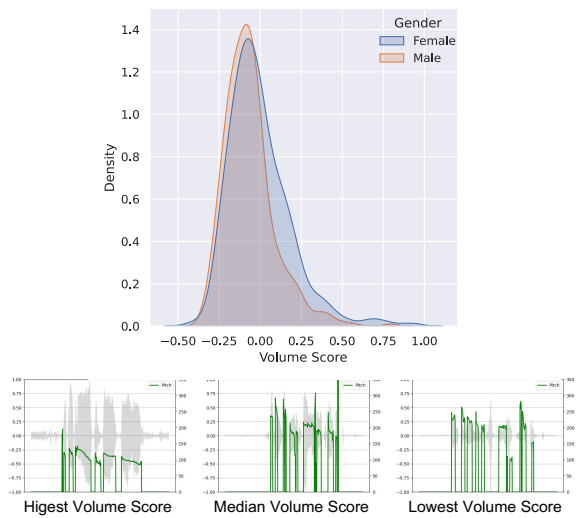

Fig. 3. **TOP:** Visualization of the probability density of pitch rank scores $r_p$. The pitch Rank module assigns $r_p$ based on the extracted pitch representation $\psi^p$. The larger the score is, the higher the pitch a speech has. We plot two densities from 900 speech samples (450 female, 450 male). We can observe that females' pitch tends to be higher than males', which is consistent with the human vocal system. **BOTTOM:** We list three speech samples with the highest, median, and lowest pitch scores. For each one, we plot the pitch curve (green) and the waveform volume (grey). We can see the pitch information is clearly different and ordered by the scores, while the overall volumes of these three samples are almost the same.

Fig. 4. **TOP:** Visualization of the probability density of volume rank scores $r_v$. The volume Rank module assigns $r_v$ based on the extracted volume representation $\psi^v$. The larger the score is, the higher the volume a speech has. We plot two densities from 900 speech samples (450 female, 450 male). We can observe that the density of females' and males' speech volumes are almost the same, both females and males can have loud or quiet voices. **BOTTOM:** We list three speech samples with the highest, median, and lowest volume scores. For each one, we plot the pitch curve (green) and the waveform volume (grey). We can see the overall volume is clearly ordered by the scores, while the pitch information of these three samples is disordered, the sample with the highest volume has a very low pitch.

*1) Visualization of Pitch and Volume Rank Scores:* In this experiment, we want to show whether the prosody encoder can capture proper pitch and volume information from the speech. Thus, we randomly pick 900 speech samples from the *unseen speakers* set, and we visualize rank scores $r^p$ and $r^v$ for pitch and volume prosody information respectively. If the pitch/volume Rank module can assign a proper score for pitch/volume information based on prosody representations (high scores for speech samples with high pitch/volume, and low scores for low pitch/volume), then it implies that the model can learn meaningful prosody representations.

In Fig. 3, on the top, we show two probability densities of pitch scores $r^p$, one for female (450 speech samples), one for male (450 speech samples). As we can see, females' speech is likely to have a higher pitch than males' speech. Such a result is consistent with the human vocal system, where a male's voice usually has a lower pitch than a female's voice. Additionally, on the bottom, we plot the pitch (the green curve) and volume (the grey waveform amplitude) for three speech samples. The sample on the left has the highest pitch score, the sample on the middle has the median pitch score, and the sample on the right has the lowest pitch score. We can observe that the pitches are ordered by the pitch score $r^p$. On the other hand, the overall volumes of these three samples are nearly the same. With such results, we can clearly see that our pitch Rank module can assign proper pitch scores $r^p$ for different speech samples, regardless how high or low of the volume, which means the pitch representation $\psi^p$, on which the score $r^p$ is based, carries useful pitch information, and unaffected by the volume information.

We depict two probability densities of volume scores $r^v$ from the volume Rank module in Fig. 4. On the top, we perform the visualization of the volume densities, one volume density is from 450 females' speech samples, the other is from 450 males' speech samples. Compared with the pitch densities (Fig. 3), there is no significant difference between females' and males' overall speech volumes. Both females and males can have loud or quiet voices. On the bottom, we also visualize the three speech samples of the highest, median, and lowest volume scores. As we can see, overall volumes drop in order of volume scores $r^v$, but volume score is unaffected by pitch, as the sample with the highest volume has very low pitch information. Such circumstances indicate that our prosody encoder is capable of recognizing volume information and extracting meaningful volume representation $\psi^v$, while the volume representations are independent of the pitch.

*2) Visualization of Speakers' Prosody Representations:* Since one main target in our work is to learn different pitch and volume representations from diverse speakers (e.g., a speaker tends to speak with a low pitch, another speaker usually speaks loudly). Thus, we visualize pitch and volume representations for unseen speakers.

We first randomly select 10 speakers from the *unseen speakers* set. To be not biased by the spoken content, we pick these speech samples with the same content "Please call

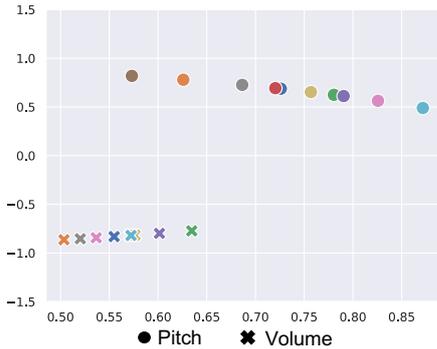

Fig. 5. Visualization of pitch and volume prosody representations for 10 unseen speakers, each color represents one speaker. We can observe: (i) Pitch and volume representations are disentangled. (ii) Our prosody encoder can extract different pitch/volume representations from distinct speakers.

Stella." By doing this, we can produce 10 pitch and 10 volume representations for each speaker.

Please keep in mind, to make our representation visualization more trustworthy, in this section (Sec. IV-C2), instead of t-SNE [33], we perform visualization on representations from an equivalent prosody encoder (Fig.1) but with 2D $\psi^p$ and $\psi^v$. In all other experiments, $\psi$ remains 512D.

We visualize pitch and volume representations in Fig. 5 and we can observe: (i) Pitch representations $\psi^p$ and volume representations $\psi^v$ are disentangled as we expected. (ii) Our $\psi^p$ and $\psi^v$ can represent prosody information for distinct speakers, which implies that our prosody encoder is able to capture prosody styles even for unseen speakers.

### D. Prosody Representation for Voice Conversion

In this section, we need to investigate whether our prosody representations can improve the zero-shot voice conversion task. Specifically, we need to validate whether the model can present the prosody information of the target speakers, and how much we can improve the speaker verification task with the addition of the prosody representations.

Therefore, we conduct experiments to evaluate the proposed model and compare it with baselines. Additionally, to see the contribution of our prosody representations for the zero-shot voice conversion task, we train 3 extra models and denote them as (i) **Ours(None)**, our VC model without the addition of any prosody representations. (ii) **Ours(P)**, our VC model with the addition of only pitch representation, and (iii) **Ours(V)**, our VC model with the addition of only volume representation.

*1) Pitch (F0) Distribution Comparison:* We conduct this experiment by following the method in [11], and show the Kullback-Leibler (KL) divergence regarding F0 distribution between the samples from the *unseen speakers* set and synthesized samples from different models. Low F0 KL divergence means the pitch information of the synthesized speech samples is close to the pitch information of the target speakers.

Specifically, we select 5 Male-to-Female (Male → Female) and 5 Female-to-Male (Female → Male) speaker pairs and use 10 audio samples for each pair. Then, we calculate log-F0 distributions of generated samples from each model and compute the KL divergence between these distributions and the distribution of original speech samples from the dataset.

We show the results in Tab. I. As we can see, our models outperform all baseline models. Furthermore, by adding pitch representation, we outperform the model without any prosody representations and the model with volume representation only. We obtain the best result when we employ both the pitch and volume representations. These findings demonstrate the importance of our pitch representation in providing valuable pitch information during VC. Additionally, **Ours(None)** also outperforms baselines, since the disentanglement methods we use (VQ, IN and CPC) potentially play crucial roles in speaker-content disentanglement, and speaker prosody information might be encoded in the speaker embedding.

*2) Volume (amplitude) Distribution Comparison:* Similar to the F0 distribution comparison, to evaluate the volume conversion performance, we compare the KL divergence between the amplitude distribution of evaluation samples and samples from our models and the baselines. The conversion of volume prosody is successful when the KL divergence between two distributions, the distribution of the volume information in the synthesized speech samples and the target speakers' volume distribution, is low.

The same 5 unseen Male-to-Female and 5 unseen Female-to-Male speaker pairs used in the F0 distribution comparison are used here. As shown in Tab. II, When the volume representation is added, the model outperforms all baselines, including our model without prosody representation and the model that just employs pitch representation. Similar to the results of the F0 Distribution comparison, the model that uses both pitch and volume representations performs the best. With these results, we can conclude that our volume representations

TABLE I
PITCH KL DIVERGENCE BETWEEN THE TARGET SPEAKERS' SAMPLES AND THE OUTPUTS OF DIFFERENT MODELS (THE LOWER, THE BETTER).

| Model | Male → Female | Female → Male |
|---|---|---|
| AGAIN [10] | 0.169 | 0.129 |
| VQVC+ [15] | 0.104 | 0.094 |
| Ours(None) | 0.099 | 0.064 |
| Ours(V) | 0.080 | 0.062 |
| Ours(P) | 0.076 | 0.049 |
| **Ours** | **0.061** | **0.035** |

TABLE II
VOLUME KL DIVERGENCE BETWEEN THE TARGET SPEAKERS' SAMPLES AND THE OUTPUTS OF DIFFERENT MODELS (THE LOWER, THE BETTER).

| Model | Male → Female | Female → Male |
|---|---|---|
| AGAIN [10] | 0.023 | 0.022 |
| VQVC+ [15] | 0.025 | 0.022 |
| Ours(None) | 0.031 | 0.030 |
| Ours(V) | 0.023 | 0.013 |
| Ours(P) | 0.023 | 0.018 |
| **Ours** | **0.019** | **0.014** |

TABLE III
SPEAKER VERIFICATION RESULTS (THE HIGHER, THE BETTER).

| Model | Cross Gender | Same Gender |
|---|---|---|
| AGAIN [10] | 0.649 | 0.671 |
| VQVC+ [15] | 0.678 | 0.692 |
| Ours(None) | 0.704 | 0.718 |
| Ours(V) | 0.706 | 0.714 |
| Ours(P) | 0.717 | 0.724 |
| **Ours** | **0.721** | **0.726** |

contain volume information that is beneficial for the VC task.

*3) Speaker Verification:* We then follow [17] and use a speaker verification tool *Resemblyzer* [3] to investigate whether the addition of prosody representation can improve speaker verification performance. *Resemblyzer* scores the speaker similarity between target speech samples and generated speech samples, on a scale of 0 to 1. The higher the score, the more similar the speakers are to the target speakers. The results are shown in Tab. III. We list the verification scores of "Cross Gender" conversion (Female to Male or Male to Female) and "Same Gender" conversion (Female to Female or Male to Male). Since "Cross Gender" conversion is more challenging and requires a more successful prosody conversion, this analysis is a robust test of our prosody representation.

As we can see from the table, our models outperform all the baselines. Furthermore, with the addition of two prosody representations, we achieve the highest score. In comparison to volume representation, we can observe that pitch representation is slightly more useful in terms of speaker verification. In conclusion, the pitch and volume extracted by our prosody encoder, especially pitch representation, are critical in improving the representation of speaker characteristics. Additionally, **Ours(None)** outperforms the two baselines, the reason might be similar to what we mentioned in Sec. IV-D1, the disentanglement methods enable encoding prosody information in the speaker embeddings.

### E. Subjective Evaluation

Subjective evaluation is another reliable method to measure the performance of zero-shot VC models. We use Amazon Mechanical Turk (MTurk) to conduct two subjective Mean Opinion Score (MOS) tests: similarity and naturalness. For the similarity test, we ask subjects to tell us whether the speakers from the synthesized and target speech are the same. For the naturalness test, we ask subjects to tell us whether the speaker from the synthesized speech speaks naturally.

We first randomly choose 10 speech samples from 10 speakers from the *unseen speakers* set. Then, we produce 10×9 = 90 conversions by generating a sample from each of the 10 speakers to each of the other 9 speakers. Since we use a vocoder PWGAN to generate waveform files from the synthesized Mel-Spectrograms, to make a fair comparison, we transform these 10 waveform files from the original dataset into spectrograms, then convert them back with the PWGAN. We denote these new waveform files as Ground Truth (GT).

[3] https://github.com/resemble-ai/Resemblyzer

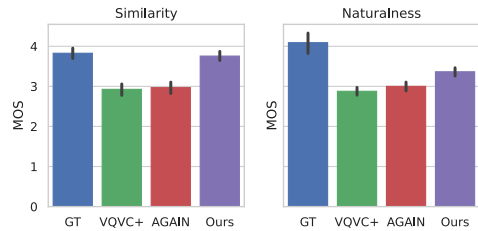

Fig. 6. Results of the MOS test. The GT (Ground Truth) comes from the dataset's samples. We set VQVC+ and AGAIN as our baselines. For the similarity test, our model performs nearly as well as GT. For the naturalness test, our model can produce more natural speech compared with the baselines.

To obtain reliable MOS results, we apply strict standards to all subjects. First, we only accept the results when subjects use headphones. We also have preliminary English-speaking questions to test whether subjects can understand English well. Trap questions are also implemented during the test to detect if subjects are just randomly picking answers. In the similarity test, the subjects are presented with pairs of utterances. Each pair has one converted sample, and one GT sample. Subjects are asked to assign a score of 1-5 on the speaker similarity: 5) Same, absolutely sure; 4) Same, sightly sure; 3) Not sure; 2) different, sightly sure; 1) different, absolutely sure. Each pair is assigned to 10 subjects. In the naturalness test, we ask subjects to assign a score: 5) Excellent; 4) Good; 3) Fair; 2) Poor; 1) Bad. Each speech sample is assigned to 10 subjects.

In Fig. 6, we show the results of the MOS tests. For the speaker similarity test, our model outperforms all baseline models and nearly achieves the same performance as GT. This implies that our extracted pitch and volume prosody can truly help represent unseen speakers for zero-shot VC as we expected. For the naturalness test, our model outperforms baselines as well and produces audios with the highest naturalness score. Thus, the MOS scores confirm that our model can successfully convert the voice among unseen speakers and synthesize speech with good naturalness.

### CONCLUSION

In this paper, we propose a novel approach to prevent the prosody leakage issue for improving zero-shot voice conversion. We introduce a prosody encoder that can learn disentangled prosody representations in a self-supervised fashion without the requirement of prosody labels or hand-crafted prosody features. We use the prosody representations as useful conditional prosody information to train a voice conversion model. To validate our approach, we perform empirical evaluations and demonstrate (i) Our prosody encoder can extract useful and meaningful disentangled pitch and volume prosody representations. (ii) The extracted prosody representations can represent prosody styles for diverse unseen speakers. (iii) Our VC benefits from the prosody representations and is able to generate natural speech samples with voices that are almost the same as the desired target speakers.